\documentclass[11pt,letter,dvips]{article}
\usepackage{epsfig}
\usepackage{wrapfig}
\usepackage{picinpar}
\makeatletter
\renewcommand{\section}{\@startsection{section}{1}{0in}
	{0.4\baselineskip}{0.1\baselineskip}{\Large\bf}}
\renewcommand{\subsection}{\@startsection{subsection}{2}{0in}
	{0.25\baselineskip}{-\baselineskip}{\large\bf}}
\renewcommand{\subsubsection}{\@startsection{subsubsection}{3}{0in}
	{0.1\baselineskip}{-\baselineskip}{\normalsize\bf}}
\makeatother
\newcommand{\icrc}{$26^{\rm th}$ ICRC\ }
\def\aap{A\&A}
\def\apj{ApJ}
\def\apss{Ap\&SS}
\def\mnras{MNRAS}

\begin{document}

%
%  Session and Paper Code:
\makeatletter\newcommand{\ps@icrc}{
\renewcommand{\@oddhead}{\slshape{OG.2.1.17}\hfil}}
\makeatother\thispagestyle{icrc}
%
%  ***INSTRUCTIONS:***  Replace `OG 9.9.9' in the command argument below
%                       with your assigned session and paper code:
%\markright{OG 9.9.9}
%

%  Title:
\begin{center}
%
%  ***INSTRUCTIONS:***  Replace `Instructions for Preparation of Manuscript'
%			with your paper's title:
{\LARGE \bf Periodicity in the TeV gamma rays and X rays from Markarian 501}
\end{center}

%  Author List:
%\begin{center}
%
%  ***INSTRUCTIONS:***  Replace authors and addresses below with your own:
%
%{\bf The Utah Seven Telescope Array Collaboration}
%\end{center}
\begin{center}
{\bf D.Nishikawa$^{1}$, S.Hayashi$^{2}$,
	N.Chamoto$^{2}$, M.Chikawa$^{3}$, Y.Hayashi$^{4}$, N.Hayashida$^{1}$,
	K.Hibino$^{5}$, H.Hirasawa$^{1}$, K.Honda$^{6}$, N.Hotta$^{7}$,
	N.Inoue$^{8}$, F.Ishikawa$^{1}$, N.Ito$^{8}$, S.Kabe$^{9}$,
	F.Kajino$^{2}$, T.Kashiwagi$^{5}$, S.Kakizawa$^{19}$, S.Kawakami$^{4}$, Y.Kawasaki$^{4}$,
     N.Kawasumi$^{6}$, H.Kitamura$^{16}$, K.Kuramochi$^{11}$, E.Kusano$^{9}$,
	E.C.Loh$^{12}$, K.Mase$^{1}$, T.Matsuyama$^{4}$, K.Mizutani$^{8}$, Y.Morizane$^{3}$,
    M.Nagano$^{18}$, J.Nishimura$^{13}$, T.Nishiyama$^{2}$,
	M.Nishizawa$^{14}$, T.Ouchi$^{1}$, H.Ohoka$^{1}$, M.Ohnishi$^{1}$, S.Osone$^{1}$,
	To.Saito$^{15}$, N.Sakaki$^{1}$, M.Sakata$^{2}$, M.Sasano$^{1}$,
    H.Shimodaira$^{1}$, A.Shiomi$^{8}$, P.Sokolsky$^{12}$, T.Takahashi$^{4}$,
	S.F.Taylor$^{12}$, M.Takeda$^{1}$, M.Teshima$^{1}$, R.Torii$^{1}$, M.Tsukiji$^{2}$,
	Y.Uchihori$^{16}$, T.Yamamoto$^{1}$, Y.Yamamoto$^{2}$, K.Yasui$^{3}$, S.Yoshida$^{1}$,
	H.Yoshii$^{17}$, and T.Yuda$^{1}$
}\\
\vspace{2.0ex}
{\it $^{1}$Institute for Cosmic Ray Research, University of Tokyo, Tokyo 188-8502, Japan\\
\it $^{2}$Department of Physics, Konan University, Kobe 658-8501, Japan\\
\it $^{3}$Department of Physics, Kinki University, Osaka 577-8502, Japan\\
\it $^{4}$Department of Physics, Osaka City University, Osaka 558-8585, Japan\\
\it $^{5}$Faculty of Engineering, Kanagawa University, Yokohama 221-8686, Japan\\
\it $^{6}$Faculty of Education, Yamanashi University, Kofu 400-8510, Japan\\
\it $^{7}$Faculty of Education, Utsunomiya University, Utsunomiya 320-8538, Japan\\
\it $^{8}$Department of Physics, Saitama University, Urawa 338-8570, Japan\\
\it $^{9}$High Energy Accelerator Research Organization (KEK), Tsukuba 305-0801, Japan\\
\it $^{10}$Department of Physics, Kobe University, Kobe 657-8501, Japan\\
\it $^{11}$Faculty of Science and Technology, Meisei University, Tokyo 191-8506, Japan\\
\it $^{12}$Department of Physics, University of Utah, Utah 84112, USA.\\
\it $^{13}$Yamagata Academy of Technology, Yamagata 993-0021, Japan\\
\it $^{14}$National Center for Science Information System, Tokyo 112-8640, Japan\\
\it $^{15}$Tokyo Metropolitan College of Aeronautical Engineering, Tokyo 116-0003, Japan\\
\it $^{16}$National Institute of Radiological Sciences, Chiba 263-8555, Japan\\
\it $^{17}$Department of Physics, Ehime University, Matsuyama 790-8577, Japan\\
\it $^{18}$Department of Applied Physics and Chemistry, Fukui University of Technology, Fukui 910-8505, Japan\\
\it $^{19}$Department of Physics, Shinshu University, Matsumoto 390-8621, Japan\\
}
\end{center}

%  Abstract:
\begin{center}
{\large \bf Abstract\\}
\end{center}
\vspace{-0.5ex}
%
%  ***INSTRUCTIONS:***  Replace text below with your own abstract:
%
Historical TeV gamma-ray flares from Markarian 501 was observed using Utah
Seven Telescope Array (7TA) from the end of March to the end of July of 1997.
The Rossi X-ray Timing Explorer All Sky Monitor (RXTE ASM) has been observing
X rays from Mrk 501 since Jan. 5, 1996.

We find evidence for periodicities of $23.9^{+2.3}_{-2.0}$ days and
$23.9^{+1.6}_{-2.9}$ days from TeV gamma-ray and X-ray light curves using 7TA
and RXTE ASM. Their false-alarm probabilities are $8.2 \times 10^{-3}$ and
$5.6 \times 10^{-5}$, respectively.
%
%  Leave this line skip in place:
\vspace{1ex}

%
%  Manuscript text:
%
%  ***INSTRUCTIONS:***  Delete the next few pages of text and enter your own.  There will
%			be a warning, `STOP DELETING TEXT!!', just before the References
%			section so that the standardized Reference heading will not be
%			accidently erased.  Within the text below is an example is given
%			of a figure placement (using `picinpar').
\section{Introduction:}
\label{intro.sec}
BL Lacertae object are considered to be a kind of Active Galactic Nuclei
(AGNs) having jet oriented to our line of the sight. Markarian 501 (Mrk 501)
is the second closest known BL Lac with red shift z = 0.034.

The big flares of Mrk 501 were detected in 1997 and varied from 0.3 up to
several times the flux from the Crab Nebula. These flares were observed
by the Whipple (Quinn et al. 1996), HEGRA (Bradbury et al. 1997), CAT (Barrau et al. 1997),
TACTIC (Bhat 1997) and the Utah Seven Telescope Array (Hayashida et al. 1998).
The CGRO EGRET has detected gamma rays (E $>$ 100 MeV) from Mrk 501 at
a significance level of $3.5 \sigma$ (Kataoka et al. 1998).
According to the All Sky Monitor (ASM) on board the {\it Rossi X-Ray Timing
Explorer} (RXTE) (Remillard and Levie, 1997), the high X-ray activity of the source
started in March 1997 and continued on October 1997.
It is found that the variation of the intensities are larger in TeV range
than X-rays and other ranges (Catanese et al. 1997, Kataoka et al. 1998, Ghisellini 1997).
Pseudo periodicity of the TeV gamma-rays in this period was reported by 7TA
(Hayashida et al. 1998) and possible suggestion of the correlation between
the TeV gamma-rays and X rays for the periodicity was reported by
HEGRA (Aharonian et al. 1998, Aharonian et al. 1999).

In this paper, we report a clear evidence of periodicities from
independent experiments which observed emissions of photons in TeV and keV
region for flares from Mrk 501.
In spite of the observations in quite different energy ranges, it shows almost
similar period of 23.9 days.

\section{Analysis:}
\label{analysis.sec}
\begin{figwindow}[1,r,%
{\mbox{\epsfig{file=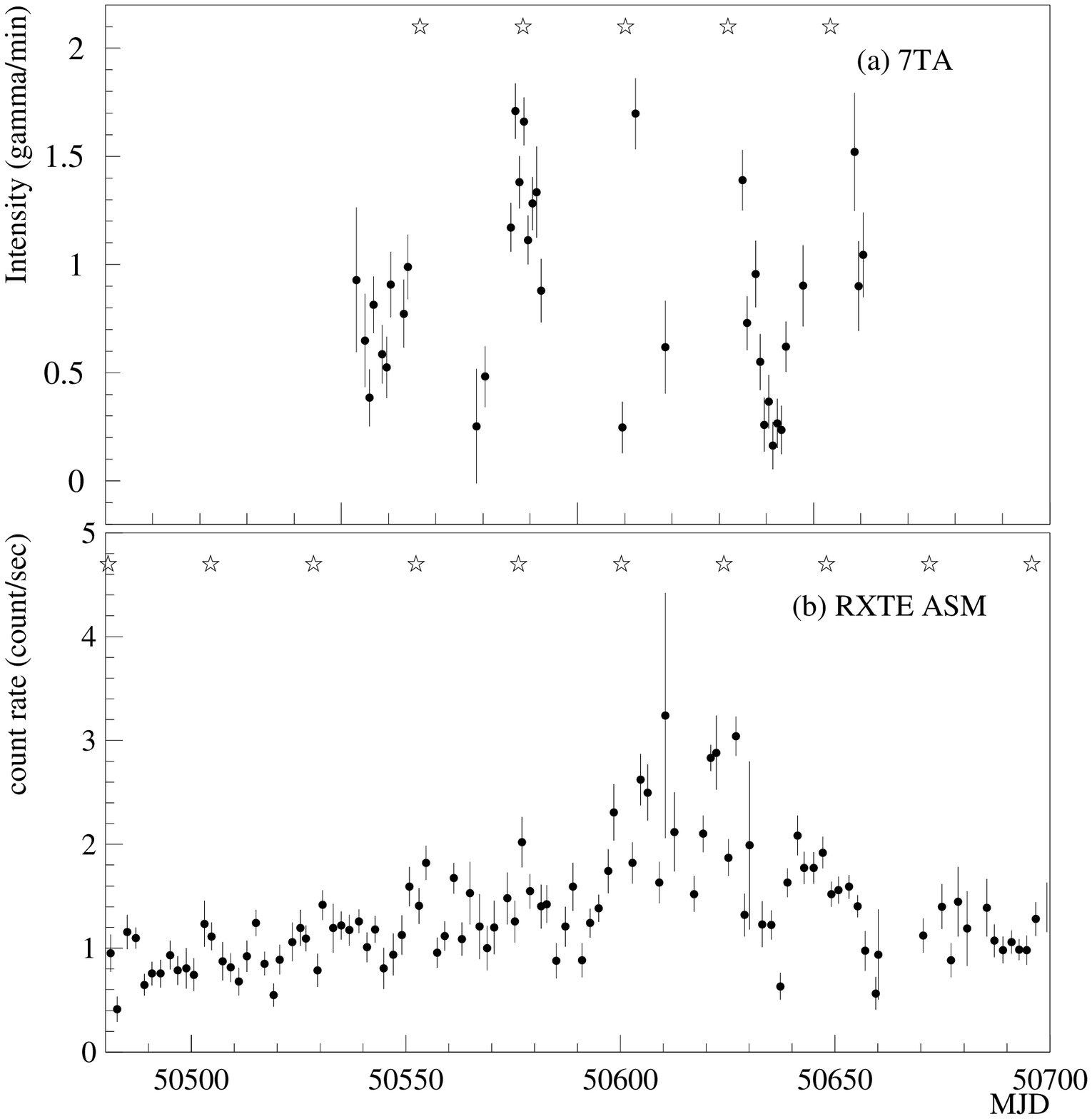,width=4.2in}}},%
{The time variation of the TeV gamma-ray and X-ray intensity
from Mrk 501 during 1997 flare (MJD 50480--50700).
Data points of the RXTE ASM light curve shows average values over 2 day
for clarity while finer datasets are used for the periodicity analysis.
Upper stars are placed at the MJD with the maximum intensity of
count rate of the fitted sine function for the obtained period.}]
We have used datasets for analysis reported by the Utah Seven Telescope Array
group for TeV gamma rays during the 1997 flares and the RXTE ASM for X rays
from MJD 50087 (Jan. 5, 1996) to 51200 (Jan. 22, 1999) for the present
analysis.

Total observation time was 105.4 hours in 47 nights during the 1997 flares.
For 7TA, we have selected data obtained under good weather condition which
are evaluation by cosmic ray event rate.
Eleven observation nights out of 47 nights are omitted in this process.

The RXTE ASM has been regularly observing bright X-ray sources
since Jan. 5, 1996. The data is publicly available over the Internet.
Dataset for RXTE ASM are selected from light curve data of SSC 1 and SSC 2.

The periodicities of the light curves were examined by the Lomb method
(Lomb 1976, Press et al. 1992) because this method can be applied to unevenly
sampled data such as observations using the air Cherenkov detector.
We can estimate the false-alarm probability from the null hypothesis by the
obtained power.
A small value for the false-alarm probability indicates a highly significant
periodic signal exists.
The errors of the period is obtained based on a simple Monte Carlo method
by generating millions of datasets with the error of the intensities
having Gaussian distribution.
\end{figwindow}

\section{Results:}
\label{results.sec}
\begin{figwindow}[1,r,%
\label{Spectra}
{\mbox{\epsfig{file=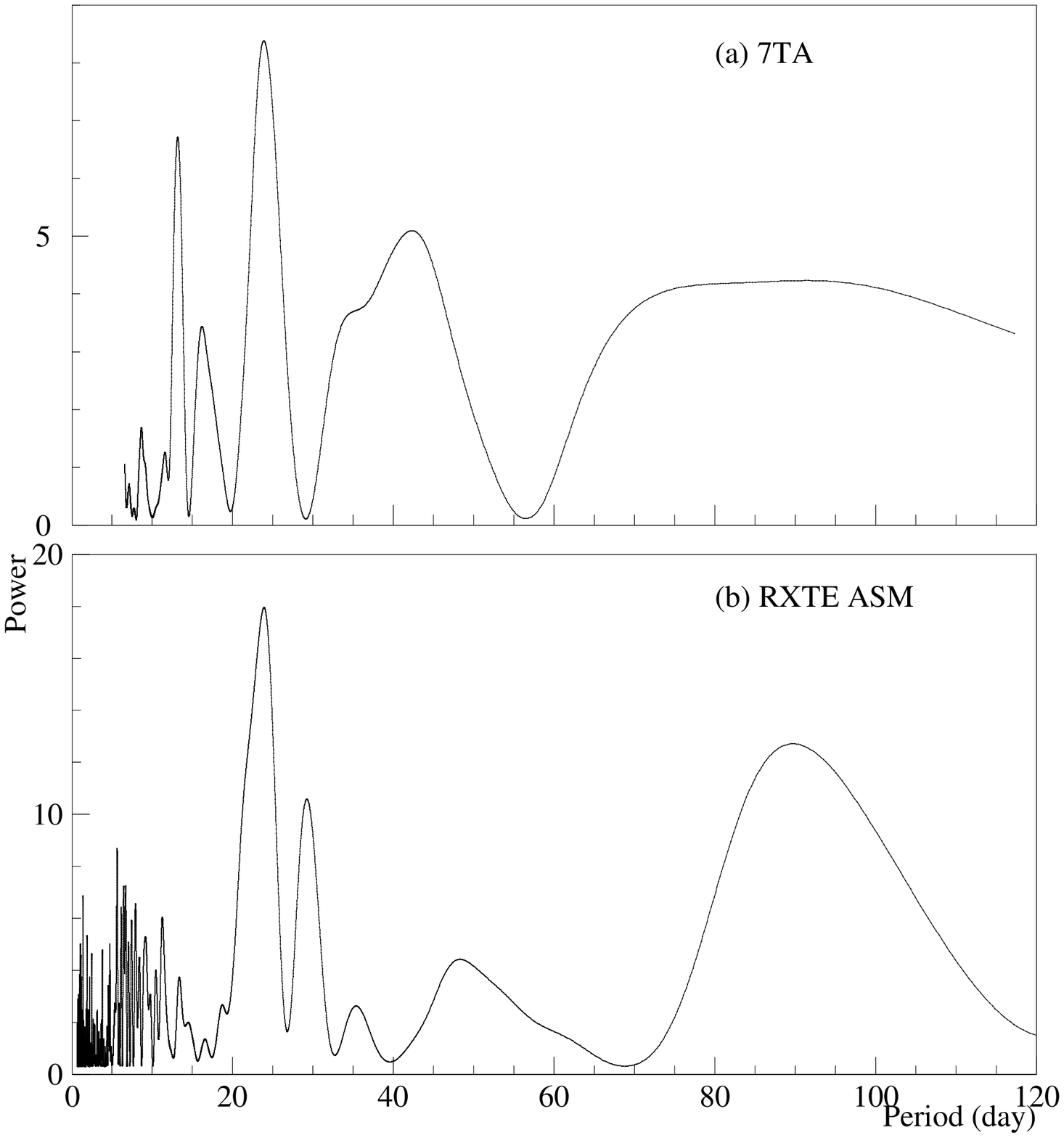,width=4.2in}}},%
{The power spectra derived from the TeV gamma-ray and X-ray light curves for
Mrk 501. The highest peaks are seen at the period of $24$ days for respective
experiments.}]
The power spectra obtained for respective experiments are shown in Figure 2.
We have generated $10^6$ datasets for 7TA and $10^3$ datasets for RXTE ASM.
The maximum peak position of the power spectra are
23.9 days (7TA) and 23.9 days (RXTE ASM).
The other peak at 13.2 days for 7TA corresponds to about a half of the 
main period of 23.9 days.

Usually, Cherenkov observations are carried out in the moon-less clear night.
Derivation of the probability could be affected by
clumping in the data sample (Horne and Baliunas 1986).
We have generated dataset which has the same gaps in the observation time 
by shuffling the observation date (in other words,
randomized the observation order of nights).
Then these dataset were analyzed in the same way as the real dataset, however,
we could not see any effect by the data clumping.
The chance probabilities are shown in Table~\ref{Prob}.
\end{figwindow}

\begin{table}
\begin{center}
\caption{Obtained periods and probabilities for the observation data of
TeV gamma-ray and X-ray emissions. 
Error of the period indicate the period change over which the power drops by
a factor of two from the peak.}
\label{Prob}
\medskip
\begin{tabular}{c c c c}
\hline
Experiment & Term [MJD] & Period [days]   & Chance probability \\
\hline
\hline
7TA        & $50538$ -- $50656$ & $23.9^{+2.3}_{-2.0}$ & $8.2 \times 10^{-3}$ \\
\hline
RXTE ASM   & $50480$ -- $50700$ & $23.9^{+1.6}_{-2.9}$ & $5.6 \times 10^{-5}$ \\
\hline
\end{tabular}
\end{center}
\end{table}

\section{Discussion:}
The power spectrum that we have computed for RXTE ASM dataset has a clear
peak at 23.9 days and the phase of light curve is consistent with the result
of TeV gamma-rays observation.
The consistent periodicities are obtained in two independent experiments.
Also the periodicity correlation between the TeV gamma-rays and X-rays
was suggested.

This periodicity may suggest that the high energy phenomena
around massive black hole(s) could be influenced by factors such as 
the rotation of the jet and the rotation of the black hole(s).
W.Bednarek and R.J.Protheroe suggested the interaction of the shock wave and
the helical structure of the jet may cause this observed type of
periodicity (Bednarek and Protheroe 1998).

\section*{Acknowledgments}
\label{acknow.sec}
The RXTE ASM results provided by the ASM/RXTE teams at MIT and at the RXTE SOF
and GOF at NASA's GSFC.
This work is supported in part by the Grants-in-Aid for Scientific Research 
(Grants \#0724102 and \#08041096) from the Ministry of Education, Science
and Culture and is also partly supported by the Promotion and Mutual Aid
Corporation for Private Schools of Japan. The authors would like to thank
the people at Dugway for the help of observations.

%
%
%
%  STOP DELETING TEXT!!
%  STOP DELETING TEXT!!
%  STOP DELETING TEXT!!
%  STOP DELETING TEXT!!
%  STOP DELETING TEXT!!
%  STOP DELETING TEXT!!
%  STOP DELETING TEXT!!
%  STOP DELETING TEXT!!
%  STOP DELETING TEXT!!
%
%
%
%  References: (DO NOT ALTER NEXT 4 LINES)
\vspace{1ex}
\begin{center}
{\Large\bf References}
\end{center}
%
%  ***INSTRUCTIONS:***  Enter your references alphabetically following the format
%			of the example citations below.
Aharonian, F. et al., 1998, submitted to \aap, (astro-ph/9808296)\\
Aharonian, F. et al., 1999, submitted to \aap, (astro-ph/9901284)\\
Aiso, S. et al. 1997a, Proc. $25^{\rm th}$ ICRC (Durban, 1997), 3, 261.\\
Aiso, S. et al. 1997b, Proc. $25^{\rm th}$ ICRC (Durban, 1997), 3, 177.\\
Aiso, S. et al. 1997c, Proc. $25^{\rm th}$ ICRC (Durban, 1997), 5, 373.\\
Barrau, A. et al., 1997, (astro-ph/9710259)\\
Bednarek, W. and Protheroe, R.J 1998, \mnras, 292, 646 (astro-ph/9802288)\\
Bednarek, W. and Protheroe, R.J 1999, submitted to \mnras, (astro-ph/9902050)\\
Bhat, C.L. 1997, Proc. Towards a Major Atomospheric Cherenkov Detector V, Kruger Park, South Africa\\
Bradbury, S.M. et al., 1997, \aap, 456, L83\\
Catanese, M. et al., 1997 \apj, 487, L143\\
Catanese, M. et al., 1998, Proc. BL Lac Phenomenon Meeting, Turku, in press (astro-ph/9810187)\\
Ghisellini, G. (astro-ph/9712286)\\
Hayashida, N. et al., 1998, \apj, 504, L71 (astro-ph/9804043)\\
Horne J.H. and Baliunas S.L., 1986, \apj, 302, 757\\
Kataoka, J. et al., 1999, \apj, 514, 138 (astro-ph/9811014)\\
Lomb, N.R., 1976, \apss, 39, 447\\
Nishiyama, T. et al, 1999, Proc. \icrc (Salt Lake City, 1999) OG 2.1.21\\
Press W.H. et al., 1992, Numerical Recipes in C, Second edition\\
Protheroe, R. J. et al., 1997, 25th ICRC (Durban, 1997) (astro-ph/9710118)\\
Quinn, J et al., 1996, \apj, 456, L83\\
Quinn, J et al., 1999, (astro-ph/9903088)\\
Remillard, R.A. and Levine, M.L, 1997, (astro-ph/9707338)\\
Samuelson, F. W., 1998, \apj, 501, L17\\
Yamamoto, T. et al, 1998, Proc. VERITAS Workshop\\
Yamamoto, T. et al, 1999, Proc. \icrc (Salt Lake City, 1999) OG 2.1.01\\
Yamamoto, T. et al, 1999, Proc. \icrc (Salt Lake City, 1999) OG 2.1.25\\
Yamamoto, T. et al, 1999, Proc. \icrc (Salt Lake City, 1999) OG 4.3.25\\

\end{document}